# Angular Patterns of Photoluminescence in Quantum Dot Spherical Superparticles Mediated by Whispering-Gallery Modes

**Yu. E. Geints**

*V.E. Zuev Institute of Atmospheric Optics, 1 Acad. Zuev square, Tomsk, 634021, Russia;*
*\*Corresponding author e-mail:* ygeints@iao.ru

**Abstract**

Quantum dot superparticles are a specific class of metamaterials created through the self-assembly of nanometer semiconductor quantum dots into organized micro-scale structures, such as microspheres. Superparticles exhibit unique optical, chemical, and electronic properties. These properties are not merely the sum of the constituent quantum dots but rather bear the signature of the collective behavior of the nanoscale building blocks. In particular, assembling an ensemble of quantum dots into a super-sphere allows them to function as a single, high-quality optical resonator. This structure efficiently confines the emission from the pump-excited quantum dots via whispering gallery modes. The emissive properties of such a superparticle resonator remain an area of active investigation. Using numerical simulation, we study the angular structure of the photoluminescence from superparticles of various sizes and architectures formed from CdS quantum dots. We show that, in general, the angular distribution of the SP emission is characterized by strong asymmetry, with a maximum in the backward direction relative to the incident pump beam. In contrast, this asymmetry is virtually absent in the forward and side-scattering directions. The excitation of resonant modes in the superparticle enhances the emission intensity and reduces the degree of its backward asymmetry. Furthermore, coating the CdS quantum dot particle with a silicon dioxide layer increases the probability of exciting field resonances in such a core-shell superparticle.

## 1. Introduction

Unlike traditional bulk semiconductors, whose electronic structure is fixed by their atomic-level bonds, light-emitting semiconductor quantum dots (QDs) represent a new class of materials for creating solid-state electronic devices [1]. The basic properties of QDs can be controllably and independently modified by optimizing material parameters and varying the chemical composition of their surface. This is important when using QDs as individual building blocks for various macro-scale devices. The unique chemical and physical properties of solid nanocrystalline QDs are utilized in the production of QD electronic circuits and optoelectronic light-emitting, sensing, and collecting devices. They are currently being actively integrated into biological and medical technologies [2, 3]. A key feature of QDs is their large surface-to-volume ratio, which creates both new opportunities and challenges when practically applying large ensembles of QDs. The dominant role of the surface and the wide range of chemical properties of materials that can be applied require continuous improvement of synthesis methods and their feedback with experimental research for specific applications in various types of devices [4].

Currently, nanoparticle synthesis methods which produce photoluminescent nanocrystalline QDs with a wide range of tunable physical properties stimulate interest in using QDs as building blocks for mesoscale assemblies known as (supra/super) particles (SPs) in spatial forms of superlattices [5], supercrystals [6], superspheres [7, 8], supraballs [9], columnar oligomers [10], etc. These nanocrystalline SPs provide functionality not observed in their constituent nanocrystals. Quantum dot superparticles (QD SPs) represent a promising platform for developing new photonic devices due to unique optical properties arising from the collective behavior of multiple nanoparticles. Typically, superparticle synthesis is carried out through the self-organization of colloidal QDs, particularly using microfluidic approaches to control the self-assembly process [11], applying emulsion technology [12], and synthesizing microparticles from colloidal solutions of quantum dots [13, 14].

In the latter case, SPs are regular spherical particles of micron-sized dimensions (less than 10 μm) with a concentration of QDs (average size of CdSe/CdS quantum dots is typically near 10 nm) of approximately $10^8$ per superparticle [15]. The spherical shape of SPs allows excitation of specific resonant configurations of the optical field — so-called high-quality Mie resonances or whispering-gallery modes (WGMs) [7, 8, 15, 16]. When the WGM frequency is precisely tuned to the corresponding emission frequency from the broad photoluminescence (PL) spectrum, the optical field of QDs inside the micron-sized SP is significantly enhanced. This has great practical potential for applications such as tunable microlasers [16, 17] and cellular microlabels [18], making these QD SPs platforms for creating next-generation microscale photonic devices. The combination of ultrafast response, high efficiency, and spectral tunability makes this technology particularly promising for practical applications in photonics and optoelectronics.

However, despite the high level of theoretical and practical development in this area, issues related to the peculiarities of the angular structure of photoemission from spherical QD SPs, the influence of particle size, and internal field resonances (WGMs) remain underexplored. In this work, we fill this gap and theoretically consider the excitation of PL emission by multiple QDs spatially organized into a spherical micron-sized particle. This leads to the formation of a PL emission within it at a shifted spectrum as a superposition of individual QD emission fields. Using numerical solutions of the Helmholtz equation for the primary (pump) and secondary (PL) electromagnetic fields by the finite element method (FEM), we model the angular structure of SP PL and determine the directions of optimal reception of emitted power.

## 2. Problem Formulation

Ignoring the nonuniform highly dispersed composition of the superparticle formed from numerous densely packed nanoscale QDs, consider a homogeneous spherical microparticle with diameter $D$ placed in air and illuminated by pump laser radiation with wavelength $\lambda_p$ (Fig. 1a). Normally, a blue laser with high photon energy in the wavelength range of 400–420 nm is used to pump QDs made of CdS/CdSe semiconductor [8, 15, 16]. To be specific, for further analysis we will use a fixed pump wavelength of $\lambda_p = 410$ nm. For the sake of simplicity, instead of considering the core-shell CdS/CdSe combination, we examine pure nanocrystalline cadmium sulfide (CdS), whose optical activity is characterized by the complex refractive index $m_{QD} = n - j\kappa$. To define the spectral dependence of $m_{QD}$ in the visible and near-IR regions, we use empirical data from Refs. [16, 19] on the optical properties of thin film composed of colloidal CdS nanocrystals and whole QDs QD superparticle.

As a result of the diffraction of the pump wave on SP, regions with enhanced optical field intensity — the «hot spots» (HSs) — are formed within sphere volume. These regions subsequently become the main sources of PL from QDs. For example, the lower graph in Fig. 1a shows the distribution of the squared modulus of the normalized pump optical field $|\mathbf{E}_p|^2$ inside a 10 µm SP (shown as an ellipse). As can be seen, due to the sufficiently high optical absorption of CdS in the UV region ($\kappa \approx 0.01$), only one HS forms near the shadowed surface inside the spherical particle. However, due to the large refractive index value ($n \approx 2.4$), the optical intensity enhancement in this region can be quite large, which creates favorable conditions for exciting radiative transitions in the quantum dot material. The corresponding distribution of field polarization sources for PL generation $|P_e|$ along the main section of the CdS SP is shown in the lower graph of Fig. 1b.

Inside and outside the particle, the interference of fields from PL sources will occur taking into account the refraction and reflection of waves at the microsphere boundary (upper figure in Fig. 1b). The resulting steady-state emission field at the shifted wavelength $\lambda_e$ is analyzed in terms of the angular distribution of its intensity in the far-field region.

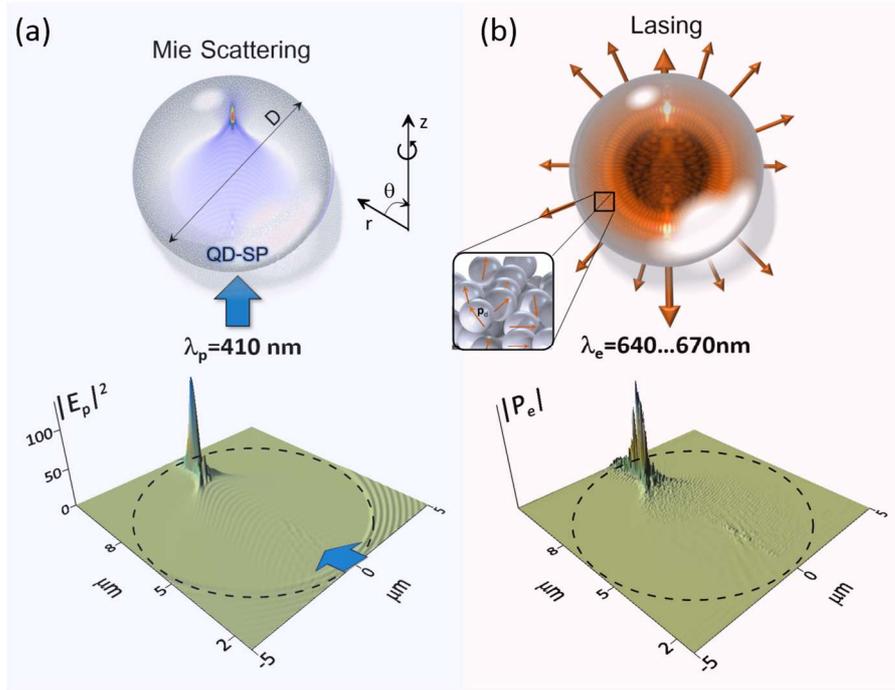

Fig. 1. (a) Artistic representation of the problem of nonlinear scattering of optical radiation by a spherical particle and the distribution of normalized optical pump intensity $|E_p|^2$ inside a 10-μm CdS SP (shown as an ellipse). (b) Illustration of the problem of QD SP photoluminescence and the distribution of PL polarization sources $|P_e|$ along the main section of the CdS SP.

Let's make several simplifications to the problem considered. First, we will consider a stationary scattering problem and artificially reduce the dimensionality of the original problem to only two spatial dimensions assuming that the distribution of optical fields is symmetric in the azimuthal direction in space. This assumption only neglects the azimuthal variation of the secondary optical field transverse to the direction of incidence of the primary radiation, while preserving the angular structure of the secondary radiation along a particular polar angle of interest.

The second simplification is considering only monochromatic radiation for both the exciting optical field and the subsequent emission from the superparticle. This means that within the discussed problem, instead of the real broadband PL of CdS QDs in the 620–680 nm range [16, 20], we can consider the emission of the entire superparticle at only single wavelength from the whole PL spectrum (at the wavelength $\lambda_e$). Particularly, $\lambda_e$ may correspond to one of the high-quality Mie resonances (whispering-gallery modes [21]) of the superparticle, which can significantly modify the angular distribution of the entire SP emission. Worthwhile noting, under the resonance conditions the quantum yield (rate) of spontaneous emission of an individual QD excited by the resonant field in the spherical microparticle is modified that is known as the Purcell effect [22]. However, we leave this effect outside the scope of this article.

Thus, within these approximations, it is assumed that each QD within the superparticle acts as a dipole that first absorbs a portion of the energy of the incident optical field $\mathbf{E}_p(\mathbf{r})$ on the microparticle at wavelength $\lambda_p$, and then spontaneously emits a photon at the shifted wavelength $\lambda_e$. Without considering non-radiative relaxation mechanisms, the rate $\gamma_d$ of spontaneous emission of a quantum dot dipole (the probability of a dipole transition) is proportional to the scalar product of the dipole moment of the corresponding emission transition $\mathbf{p}_d$ and the local field strength $\mathbf{E}_p$ raised to the power of $m$ [23, 24]:

$$\gamma_d^{(m)} \propto \alpha_p^{(m)} \left| \mathbf{p}_d \cdot \left[ \mathbf{E}_p(\mathbf{r}) \right]^m \right|^2, \tag{1}$$

where $m$ is the number of simultaneously absorbed optical photons, and $\alpha_p^{(m)}$ is the $m$-photon spectral absorption coefficient of the material at wavelength $\lambda_p$.

In describing spontaneous emission from a spherical particle, the following inhomogeneous wave equation is solved within the stationary problem formulation [25]:

$$\left( k_0^2 \varepsilon_1 \right)^{-1} \nabla \times \left( \nabla \times \mathbf{E}_e(\mathbf{r}) \right) - \mathbf{E}_e(\mathbf{r}) = \mathbf{P}_e \left\{ \mathbf{E}_p(\mathbf{r}); \tilde{\mathbf{r}}_1 \right\}, \tag{2}$$

where $k_0$ is the wave number in free space, $\varepsilon_1 = m_{QD}^2$, and the polarization source $\mathbf{P}_e$ is some random function on the coordinates of radiating dipole $\tilde{\mathbf{r}}_1 = (\tilde{r}, 0, \tilde{z})$:

$$\mathbf{P}_e(\mathbf{r}; \tilde{\mathbf{r}}_1) \propto \alpha_p^{(m)} \left( \mathbf{E}_p \right)^{2m-1} \left( \tilde{\mathbf{n}}_d \cdot \mathbf{n}_r \right), \tag{3}$$

where $\mathbf{n}_r = \mathbf{r}/|\mathbf{r}|$, and $\tilde{\mathbf{n}}_d \equiv \mathbf{n}_d(\tilde{\mathbf{r}}_1)$ is a randomly directed unit vector of the molecular dipole moment ($\mathbf{p}_d = |\mathbf{p}_d| \mathbf{n}_d$). Then, the rate of energy exchange between the incident UV radiation ($m = 1$) and the QD dipole, $dW/dt = \left( \mathbf{E}_p^* \cdot \mathbf{P}_e \right)$, is proportional to the pump intensity: $dW/dt \propto I_p$, where $I_p = E_p^2$ is the optical field intensity. For the case of two-photon absorption ($m = 2$), we obtain a dependence on the square of the pump intensity: $dW/dt \propto I_p^2$.

The numerical simulation of SP photoluminescence is carried out by numerically solving Eqs. (2)-(3) using the finite element method, implemented in the COMSOL Multiphysics software. Since the photon emission of secondary radiation by a QD typically occurs spontaneously, in the numerical model the radiation dipole sources at each point in space in general are uncorrelated both in phase and direction. In the following, within the stationary approximation, we only consider the random direction of dipole moments $\mathbf{p}_d$ by specifying a random polar angle of the dipole $\theta \to \tilde{\theta}(\mathbf{r})$ in the range $[0, 2\pi]$ in the ($r$-$z$) plane, as shown in Fig. 1b. Then, at each point with coordinates $\mathbf{r}$ inside the particle, the unit vector of the dipole moment can be written as a function of only one random parameter: $\tilde{\mathbf{n}}_d = \hat{\mathbf{r}}\sin\tilde{\theta} + \hat{\mathbf{z}}\cos\tilde{\theta}$, where $\hat{\mathbf{r}}$ and $\hat{\mathbf{z}}$ are unit vectors along the corresponding coordinate axes.

In the calculations, statistical averaging is performed by specifying multiple variants (usually one hundred) of the spatial distribution of dipole moments of radiating QDs with random directions $\tilde{\mathbf{n}}_d$, followed by solving problem (2)-(3). After a series of calculations, all computed fields of secondary radiation are averaged over space and then the resulting averaged field distribution is used to calculate the PL angular pattern in the far-field diffraction zone ($r \gg D$) using the Stratton-Chu formulas [26]. This formalism is based on the assumption that the Green's function for the vector Helmholtz equation (2) in the far-field is known, and the propagation medium is homogeneous in this region. Then, the electric field $\mathbf{E}_{far}$ in the far-field zone at any point $\mathbf{r}$ on the surface of some hypothetical sphere $S$ with radius $r$ and normal $\mathbf{n}$, enclosing the emitting "antenna" SP, is expressed through the integral of the corresponding near-field as:

$$\mathbf{E}_{far}(\mathbf{r}) = \frac{jk_0\mathbf{r}}{4\pi} \times \int_S \left[ \mathbf{n} \times \mathbf{E}_e(\mathbf{r}_1) - \eta\mathbf{r} \times (\mathbf{n} \times \mathbf{H}_e(\mathbf{r}_1)) \right] \exp(jk_0\mathbf{r}_1 \cdot \mathbf{r}) dS, \qquad (4)$$

where $\eta = \sqrt{\mu/\varepsilon}$ is the medium impedance, and $j$ is the imaginary unit. Using Eq. (4) it is possible to calculate the angular distribution of PL emission intensity $I(\theta) = |\mathbf{E}_{far}|^2$ from a QD SP based on the results of calculating the electromagnetic fields $\mathbf{E}_e$ and $\mathbf{H}_e$ directly near the particle.

## 3. Angular Patterns of PL from Superparticles

First, consider the spatial structure of QD SP emission in the near-field region, i.e., directly inside the microparticle and at a small ($\sim\lambda_e$) distance from its outer surface. Clearly, the spatial emission pattern will be determined by the configuration of nonlinear polarization sources within the particle, formed under the action of the pump optical field. As the analysis of the pump wave intensity profiles $I_p$ (Fig. 1a) shows, if the SP is sufficiently (optically) large, only one "hot spot" always forms inside the particle with an absolute intensity maximum located near the shadowed hemisphere rim. This "hot spot" serves as the polarization source $\mathbf{P}_e$ for the PL field produced by the quantum dots. This characteristic of the spatial arrangement of the pumping field within the SP is a result of the intense absorption in the QD material (CdS). In this case, optical rays hitting the center of the superparticle significantly lose their intensity, and the principal HS is formed by lateral grazing rays that are focused by the curved surface of the supersphere.

The maximum optical pump intensity achieved in the shadow HS of superparticles of various sizes is shown in Fig. 2a. The same figure shows the dependence of the absorption efficiency $\sigma_{ab}$ of pump radiation by the entire SP, defined as the ratio of the total optical power absorbed in the particle to the power of radiation incident on its cross-section. Here and in what follows, due to the geometry of the model used, the initial radiation is considered to be circularly polarized relative to the incidence direction.

As can be seen, with increasing SP size, first there is an increase in optical intensity of pump radiation in the internal focusing region, followed by saturation at $D \approx 6$ μm and even a decrease in this value for even larger particles. At the same time, according to Mie theory [27], the absorption cross-section $\sigma_{ab}$ of the SP demonstrates a monotonic increase with increasing sphere size, which then transits to saturation near the limiting value $\sigma_{ab}(D \to \infty) = 1$. This means that for the most efficient conversion of pump energy into PL field of QDs, it is not necessary to produce excessively large SPs, and spheres with sizes of several microns are quite sufficient.

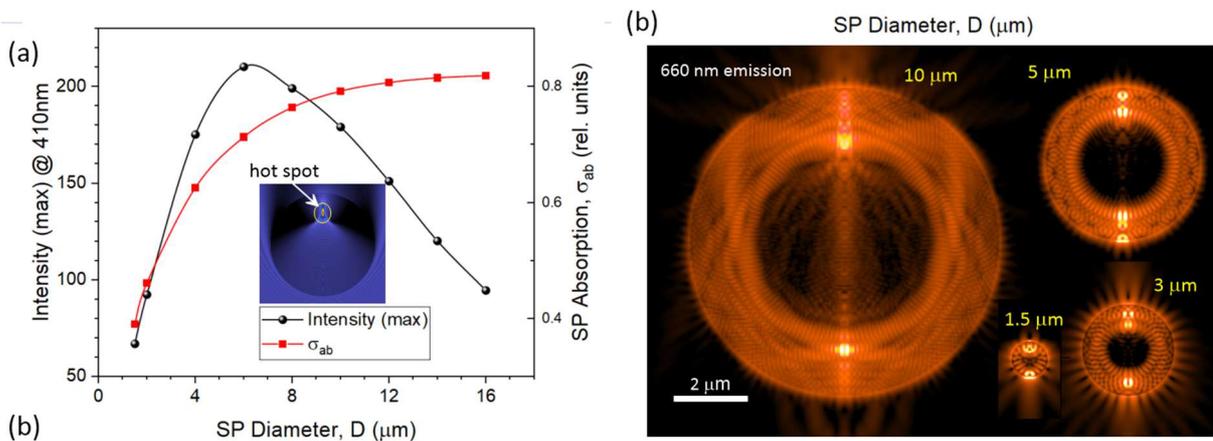

Fig. 2. (a) The maximum pump intensity (410 nm) in the "hot spot" and the normalized absorption cross section for QD SPs of various diameters. (b) Spatial distribution of PL at $\lambda_e = 660$ nm in the vicinity of SP of various sizes (yellow labels) when pumped with UV laser (410 nm, bottom-top).

The pattern of QD SP photoluminescence of various sizes is presented in Fig. 2b as a spatial distribution of intensity $|\mathbf{E}_e|^2$ along the principal cross-section of the particle. Since the source of QD photoluminescence is nonlinear polarization $P_e$, the emission inside the particles also concentrates mainly near the surface, although it has a more complex spatial structure due to interference during multiple reflections from the inner spherical surface.

These successive reflections are clearly visible for 10 μm SP as the ballistic trajectories of optical rays emitted by the principal HS in the shadowed hemisphere of the particle. The result of such multiple ray reflection is the formation of a spherical caustic in the central region and an additional HS in the illuminated part of the sphere. Meanwhile, the central region of the SP visually remains dark against the background of bright peripheral regions, which corresponds well to experimentally recorded images of luminescent CdS/CdSe SPs of micron sizes (dark-field microscopy) [8, 16, 20]. The presence of two main emission sources causes the specific structure of SP photoluminescence in the near-field region with intensity maxima in the forward and backward directions relative to the pump radiation.

Interestingly, for particles with diameters of 5 μm and 3 μm, the PL field structure has a clearly defined resonant character with the presence of equidistant nodes and antinodes of optical intensity along the spherical particle rim. This indicates the excitation of Mie resonances in the form of WGMs [28] at this PL emission wavelength, when the approximate condition [29] is satisfied: $D/\lambda_e \cong (l + \frac{1}{2})/\pi|m_{QD}|$, where $l$ is the azimuthal mode index specifying the number of optical field maxima in the angular direction (on the spherical semicircle).

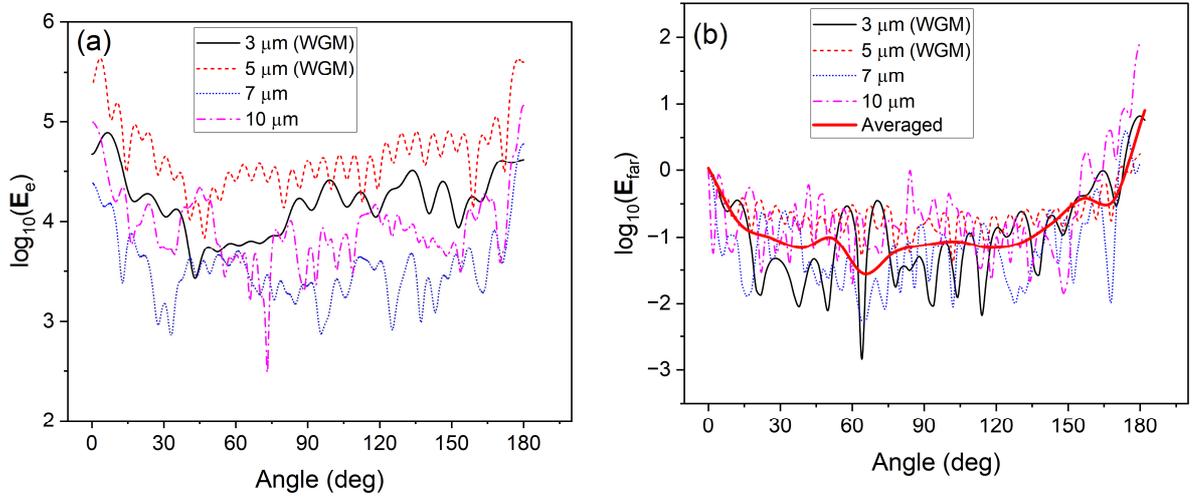

Fig. 3. Angular structure of SP photoluminescence in the near- (a) and far-field regions (b).

In resonance, the spatial structure of the internal PL field becomes almost symmetric relative to the particle equator. This leads to corresponding symmetrization of the angular distribution of radiation optical field near the microparticle, as shown in Fig. 3a. This figure presents the angular dependence of the photoluminescence from SPs of various sizes in the near-field. The pump radiation exciting QD emission is incident in the direction of 180°→0° angular axis. Evidently, due to the excitation of WGMs in particles with diameters of 3 μm and 5 μm, the near-field PL generally possesses the highest intensity and is characterized by a higher degree of symmetry relative to the forward-backward propagation direction. In a 10 μm SP, a certain symmetry in the angular distribution of the emission field is also observed, but it is not as pronounced as in the case of 5 μm SP, since the internal field is not precisely tuned to the resonance.

Now, consider the far-field patterns of laser-induced PL from SPs. As previously mentioned, the angular distribution of particle PL intensity is calculated using integral expressions (4), obtained under the assumption that in the Fraunhofer diffraction zone, i.e., sufficiently far from the microparticle, only the scattered field exists, and the initial optical field equals to zero. This allows transitioning from a cylindrical to a polar coordinate system and considering the dependence of field vectors only along the polar angle.

The results of calculating the PL phase function $E_{far}(\varphi)$ for SP of various sizes are shown in Fig. 3b. For better understanding, all values are normalized to the field amplitude at $\varphi = 0°$. Recall, that presented angular structure of SP emission is statistically averaged over approximately one hundred runs of the random configuration of QD radiating dipoles at each point within the superparticle. The standard deviation of PL emission intensity for each angular direction depends on the conditions for exciting Mie resonances and is about 20% for the non-resonant case and 8% for PL emission under WGM conditions.

From examining Fig. 3b it follows that the symmetry of PL angular distribution near the spherical particle leads to a similar symmetric structure in the far field. This mostly applies to the 5-μm SP, in which a WGM resonance is excited at a wavelength of 660 nm on the **TE**$_{35,4}$ mode with 35 azimuthal and 4 radial maxima of optical intensity and a quality factor $Q = 437$, as shown in Figs. 4a and b. In smaller particle with $D = 3$ μm (Fig. 4a), there is also an internal optical resonance at 660 nm on the **TE**$_{16,4}$ WGM.

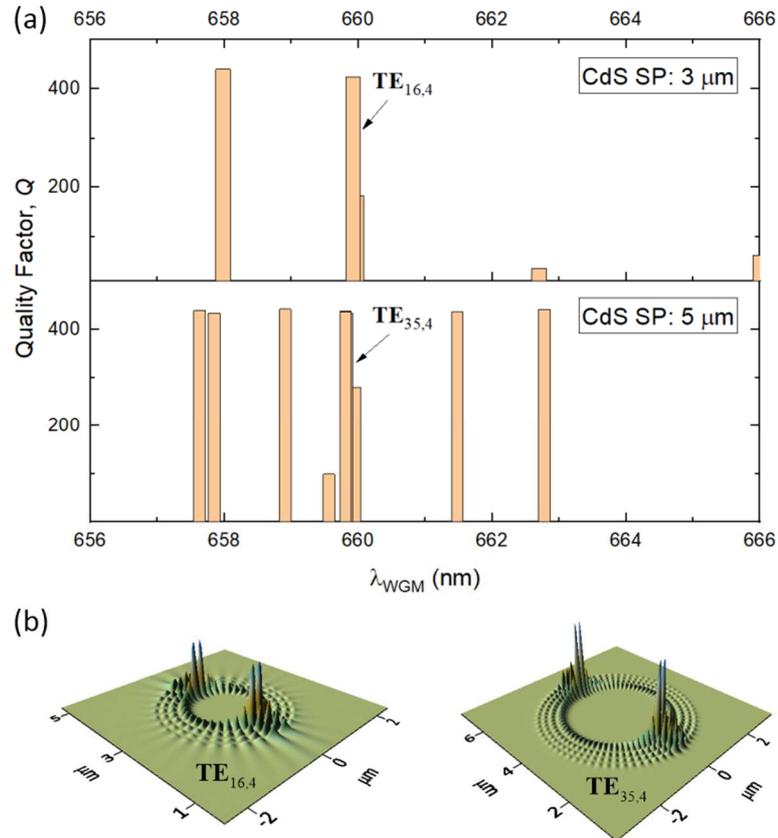

Fig. 4. (a) WGMs in PL spectrum of QD SPs with different sizes. (b) The resonant structure of the optical field for **TE**$_{16,4}$ и **TE**$_{35,4}$ WGMs.

Despite the relatively high refractive index of CdS ($n = 2.24$), low values of WGM quality factors are exclusively due to the rather high optical absorption of the QD material in the visible spectrum region, where $Q \simeq Q_{abs} = \mathrm{Re}(m_{QD})/2\,\mathrm{Im}(m_{QD})$ [30, 31], while $m_{QD} = 2.24 - j\,0.0026$ at $\lambda = 660$ nm.

Let's consider an important quantitative characteristic of the angular distribution of particle emission, namely, the angular directivity of emission in the backward/forward $D_{SP}^{back}$ and side/forward directions $D_{SP}^{side}$, which we will define as follows:

$$D_{SP}^{back} = 10 \cdot \log\left\{ \int_{3\pi/4}^{\pi} E_{far}^2(\varphi)\,d\varphi \bigg/ \int_0^{\pi/4} E_{far}^2(\varphi)\,d\varphi \right\}, \qquad (5)$$

$$D_{SP}^{side} = 10 \cdot \log\left\{ \int_{\pi/4}^{\pi/2} E_{far}^2(\varphi)\,d\varphi \bigg/ \int_0^{\pi/4} E_{far}^2(\varphi)\,d\varphi \right\}, \qquad (6)$$

These parameters as a function of particle size are shown in Fig. 5. It is clearly seen that the SP luminescence mainly occurs in the direction opposite to the incidence of the pump radiation, i.e. the particle glows brightest when viewed from the side of the hemisphere illuminated by the pump. This effect of backward enhanced emission was previously observed for the multiphoton-excited fluorescence of dyed droplets and is related to the principle of reciprocity of optical fields inside and outside a spherical scatterer [24, 32]. When resonances are excited in the SP, PL emission directivity to the backward direction decreases markedly. Generally, in the lateral direction, SPs glow approximately with similar intensity (in integral terms) as in the forward direction, except for particles of small sizes, $D < 3$ μm, when the dipole nature of the QDs emission begins to manifest itself with the appearance of lateral lobes in the angular distribution.

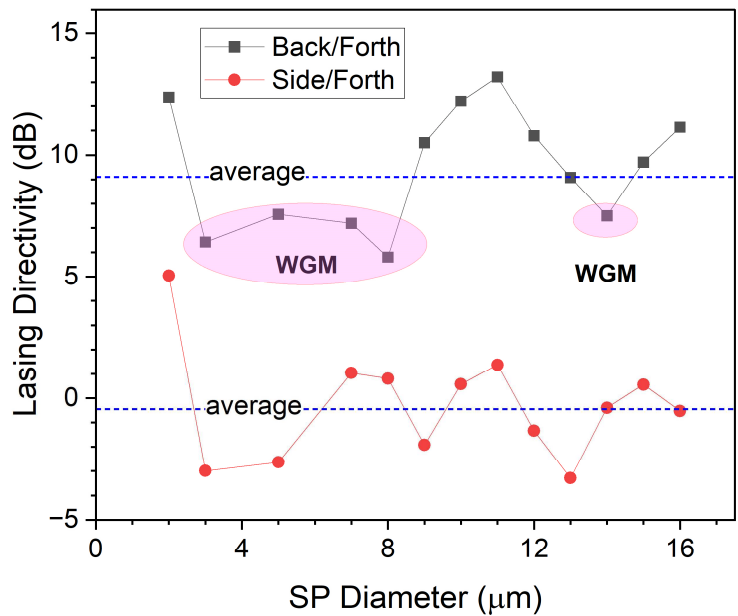

Fig. 5. PL directivity of SPs depending on their diameter.

Clearly, in general, the PL angular distributions in the far-field for all considered particle sizes exhibit similarity, demonstrating a parabolic type of dependence with two maxima in the forward and backward directions. Therefore, it is instructive to average all these phase functions to obtain one averaged PL angular dependence, which is shown in Fig. 3b as a bold red curve. It is expected that this averaged dependence can most accurately characterize the angular structure of a nonresonant PL of any SP (micron size), since in real conditions achieving resonant WGM excitation in a spherically imperfect composite SP is quite a difficult task. Such an averaged PL angular pattern has predominantly backward emission directivity with parameters $D_{SP}^{back}$ = 9.1 dB and $D_{SP}^{side}$ = -0.5 dB which are marked by dashed lines in Fig. 5.

## 4. Two-Photon Absorption in QD SP Fluorescence

It was previously noted that the main source of the photoluminescence inside the SP is a single "hot spot" in the shadow part of the microsphere formed by optical rays reflected only once from the inner surface of the particle as illustrated by the geometrical sketch in Fig. 6b. This is due to the relatively high optical absorption of CdS in the UV range, as shown in Fig. 6a, which presents the spectral absorption of QD material.

Indeed, at the pump wavelength of 410 nm, CdS absorption is almost $10^{-2}$, which considering the high refractive index of CdS at this wavelength ($n$ = 2.37) gives a single-photon absorption (SPA) inverse length, $\alpha = 4\pi n\kappa/\lambda$, inside a SP of approximately 0.6 μm$^{-1}$. In other words, in microparticles larger than $D \sim 2/\alpha$, optical rays after reflection from the rear surface will not reach the front hemisphere with sufficient intensity to form a second polarization source for QD photoluminescence.

However, the absorption of CdS sharply decreases when shifting to the near-IR spectral region, for example, to the operating range of Ti:Sapphire lasers (750–950 nm). This opens up prospects for using a QD SP as an optical resonator not only at PL wavelength but also at the pump wavelength. Historically, this situation is called double resonance scattering [33, 34], when there is a significant increase in the intensity of Raman scattering or lasing from a liquid microsphere, provided that both incident and nonlinearly scattered (fluorescent) optical waves become resonant with some eigenmodes of a micron-sized particle.

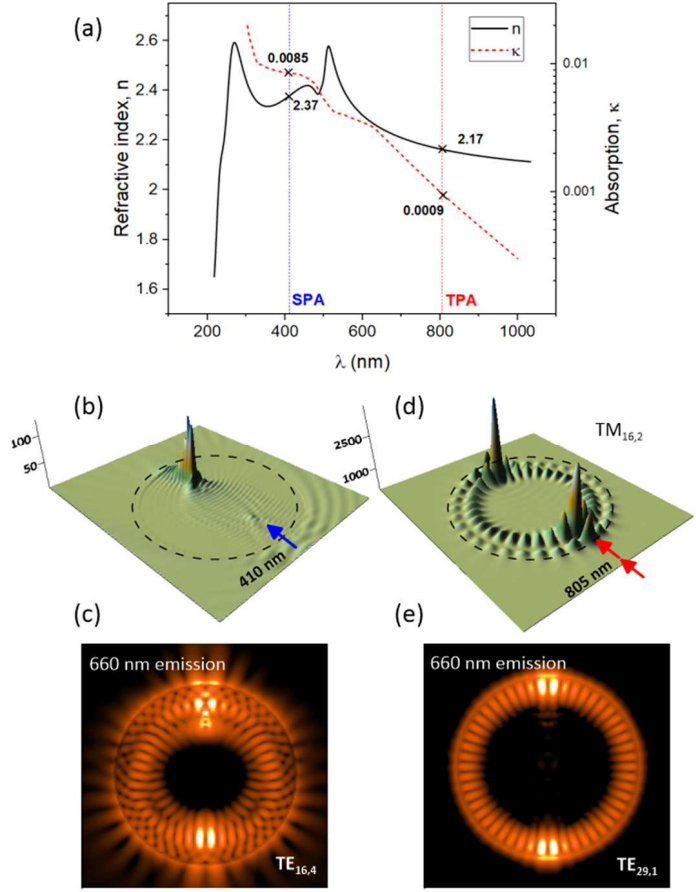

Fig. 6. Double resonance excitation via TPA in QD SP ($D = 3$ μm) by an IR pump radiation. (a) Spectral dependences of refractive index ($n$) and optical absorption (κ) of CdS QD. (b, d) Distribution of pump intensity at 410 nm (b) and 805 nm (d) inside the SP. (c, e) Patterns of PL intensity during single-photon (c) and two-photon absorption (e).

It is clear that in the case of IR pumping, the large band gap of CDs (~2.4 eV) must be overcome by optical radiation with low photon energy (~1.5 eV), which is possible only when using a high-intensity laser pulse sufficient to realize two-photon absorption (TPA). in the QD material. Currently, Ti:Sapphire lasers provide this capability by generating femtosecond optical pulses with typical peak intensities of several TW/cm² and overcoming the threshold for developing second-order nonlinearity [35] and even third-order nonlinearity [36] in CdS QDs.

We simulate such double-resonance PL in a 3 μm SP by tuning the incident radiation to one of the WGMs of the microsphere, specifically the **TM**$_{16,2}$ mode with a resonant wavelength of $\lambda_{WGM} = 804.83$ nm. In the calculations, it is assumed that when implementing TPA in SP material {$m = 2$ in Eq. (3)}, QDs emission occurs at a wavelength of $\lambda_e = 660$ nm due to the formation of polarization sources ~ $I_p^2$ in the spatial structure of the WGM, as shown in Fig. 6d. The spatial distributions of SP emission during SPA and TPA regimes inside the particle are shown in panels (c) and (e) of this figure.

As seen, in the case of double resonance PL the specific resonant distribution of the pump field, which is not deeply penetrating into the particle, promotes excitation of a high-quality "shallow" **TE**$_{29,1}$ WGM with a quality factor $Q \approx 500$ at the PL wavelength compared to the less-quality "deep" **TE**$_{16,4}$ mode shown in Fig. 6c. Obviously, this will also lead to a change in the total PL intensity of SP, although within the framework of this theoretical model it is difficult to estimate this effect.

## 5. Core-shell QD Superparticles: PL Angular Patterns

Recall Fig. 2a for the pumping intensity distribution during single-photon absorption in the QD material. As previously noted, the main source of nonlinear polarization for the PL inside the SP is the region of maximum intensity enhancement in the shadow hemisphere. This maximum is formed as a result of refraction of the incident field at the spherical particle surface and subsequent reflection from the inner surface, which acts as a semi-transparent mirror due to high refractive index contrast $m_{QD}/n_{air} > 1$. The relatively strong optical absorption of CdS prevents the intensity inside the particle from reaching high values. However, this parameter can be controlled by adding a UV-transparent shell with good focusing properties, for example, made of commonly-used silica (SiO$_2$) with a refractive index $n = 1.47$ and negligible optical absorption ($\alpha < 10^{-4}$ [37]).

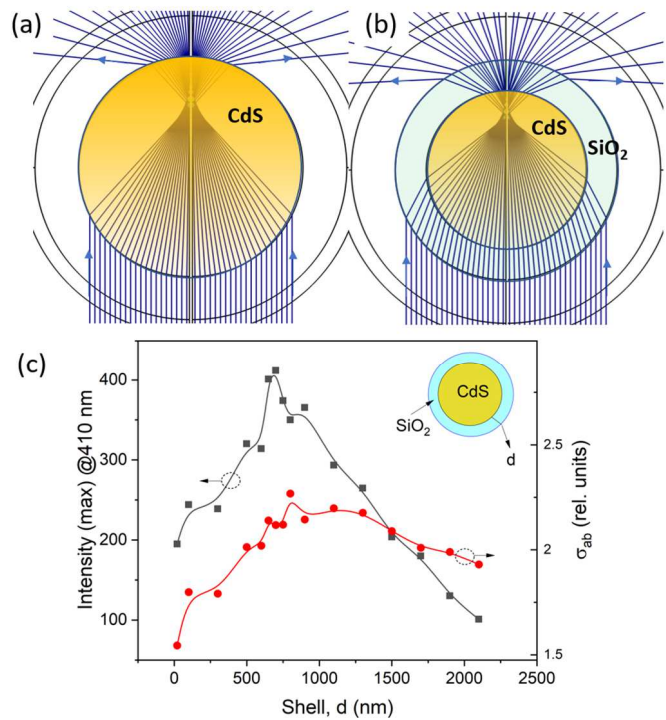

Fig. 7 (a, b) Geometric optics of SP. Ray tracing inside a (a) solid and (b) core-shell SP. (c) Change in maximum pumping intensity and normalized absorption cross-section in a 5 μm CdS-SiO$_2$ SP depending on the thickness $d$ of the silica shell.

The principle of operation of such a focusing shell is illustrated in Figs. 7a and b, which show the trajectories of geometric rays been incident on a solid and core-shell particle. As can be seen, the non-absorbing silica shell additionally focuses optical rays, which subsequently converge in the shadow "hot spot" at large angles. This reduces the spatial extent of the focal region and moves it away from the particle center (in relative value) while simultaneously increasing the optical field intensity. In addition, the less contrasting optical shell reduces optical reflection from SP illuminated surface and acts as an anti-reflective coating.

Meanwhile, a shell that is too thick can shift the internal focus outside the absorbing CdS core, reducing PL output, while a shell that is too thin will not provide the desired effect. As a result, there exists an optimal ratio of core-to-shell sizes $D/d$ at which the internal pumping intensity and overall particle absorption are maximized. For a 5 μm CdS SP, as shown in Fig. 7c, this optimum occurs when the shell thickness is approximately between 600 and 800 nm.

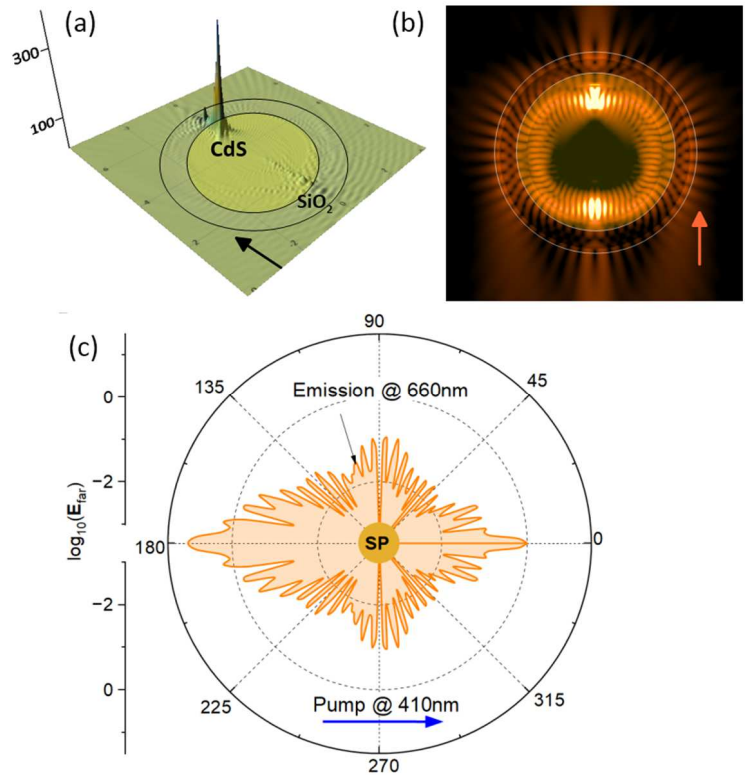

Fig. 8. PL of a CdS-SiO$_2$ SP with core-shell parameter $D/d$ = 5000/700 nm. (a) Pump intensity distribution; (b) SP photoluminescence in the near-field. (c) Angular structure of photoluminescence in the far-field region.

PL field structure is shown in Figs. 8a-c calculated for a CdS-SiO$_2$ SP with the core-shell ratio $D/d$ = 5000/700 nm. In addition to the significantly higher field intensity in the shadow "hot spot", a specific feature of the near-field PL distribution for such a particle is the excitation of a higher-quality whispering gallery mode **TE$_{28,1}$** with azimuthally-symmetric structure. Compared to the CDs solid particle of the same size ($D = 5$ μm) shown in Fig. 2b, the WGM excited in the shell-type SP is located closer to the surface of the SP absorbing core and is scattered by the outer spherical silica shell. This effect looks like a wide scattering of optical rays refracted in the shell in Fig. 8b and leads to a more symmetric angular structure of PL in the far-field (Fig. 8c). The evaluation of PL angular directivity for such core-shell SP using Eqs. (5) and (6) gives the following values $D_{SP}^{back}$ = 10.1 dB and $D_{SP}^{side}$ = -1.6 dB.

## 6. Conclusions

To summarize, in this work we theoretically study the angular characteristics of PL emission from an ensemble of CdS QDs spatially organized into a single spherical SP pumped by optical radiation with the possibility of stimulating two-photon absorption and exciting resonant WGMs within microparticles. Using numerical simulation of the random distribution of nonlinear polarization sources within the SP volume, we calculate the angular structure of the PL in the near- and far-field diffraction and establish the dependence of PL emission directivity on SP size and structural type (solid/core-shell).

We show, that in accordance with the reciprocity principle, QD SP emission occurs primarily in the direction opposite to the UV pump radiation and to a lesser extent in the direction of its incidence. At the same time, SP PL in the lateral direction is always less intense but can be enhanced in the case of resonance with the particle WGMs. The averaged angular directivity of SP emission in the backward/forward and side/forward directions is approximately 9.1 dB and -0.5 dB, respectively. The presence of a non-absorbing spherical shell over a CdS core leads to increased pump and PL intensity and a more symmetric angular pattern of photoluminescence in the far-field due to additional concentration and scattering of incident pump wave by transparent spherical shell.

QD SPs represent a significant advancement in the development of nanomaterials. By transitioning from isolated QDs to organized metastructures researchers can utilize new collective properties such as improved light confinement and ultrafast optical switching. This makes QD SPs promising devices for applications in photonics, sensing, and quantum technologies. Currently, the main application areas of QD SPs are advanced photonics and multi-analyte chemical sensing. We expect that the results of this work will be particularly useful in developing a theoretical model of laser-induced stimulated photoluminescence (lasing) of QD agglomerates assembled in different-shapes SPs for various academic and technological applications.

**Acknowledgements**. Ministry of Science and Higher Education of Russian Federation.

**Disclosures**. The authors have no relevant financial or non-financial interests to disclose.

**Author Contributions.** All authors contributed equally to the study conception and design. All authors read and approved the final manuscript.

**Data availability**. The datasets generated during and/or analyzed during the current study are available from the corresponding author on reasonable request.